\documentstyle[preprint,aps,floats]{revtex} 
 
\begin{document} 
\draft

\input epsf

\title{Mesoscopic conductance fluctuations in dirty quantum dots
with single channel leads} 
 
\author{Edward McCann and Igor V.\ Lerner} 
\address{School of Physics and Space Research, University of  
Birmingham, 
Birmingham~B15~2TT, United Kingdom 
} 
\maketitle 
 
\begin{abstract} 
\small\baselineskip=18pt 
We consider a distribution of conductance fluctuations in 
quantum dots with single channel leads and continuous level
spectra and we demonstrate that it has  a distinctly non-Gaussian
shape and  strong dependence
on time-reversal symmetry, in contrast to an almost  Gaussian distribution
of conductances in a disordered metallic sample connected to a reservoir by
broad multi-channel leads. In the absence of time-reversal symmetry,
our results obtained within the diagrammatic approach coincide with those
derived within non-perturbative techniques.
In addition, we show that
the distribution has lognormal tails for weak disorder, similar to
the case of broad leads, and that it becomes almost
lognormal as the amount of disorder is increased
towards the Anderson transition.
\end{abstract} 

\def \abt#1{\langle #1 \rangle} 
\def \aabt#1{\langle {\langle #1 \rangle} \rangle} 
\def \gplus#1{{\cal G}^+ \left( #1 \right) }
\def \gminus#1{{\cal G}^- \left( #1 \right) }
\def \diff#1{D \left( #1 \right) }

\pacs{PACS numbers: 72.15.-v, 73.20.Dx, 73.20.Fz} 
 
 \baselineskip=19pt 
 
Recently it has been shown that the conductance of clean quantum dots with 
point-like external contacts (lead width $w \approx \hbar  k_F^{-1}$, where 
$\hbar  k_F^{-1}$ is the Fermi wavelength) have non-Gaussian distribution 
functions \cite{Pri:93,Jal:94}.
 Weak transmission through the contacts means that the electrons typically 
spend more time in the system than that required to cross it so that the 
energy level broadening due to inelastic processes in the dot $\gamma \ll  
E_c$ where $ E_c \approx \hbar D/L^2$ is the Thouless energy.
 This inequality corresponds to the zero mode regime which allows the use 
of non-perturbative techniques including random matrix theory \cite{Meh:91} 
and the zero dimensional supersymmetric $\sigma$ model \cite{Efe:83}.
In contrast, it is known that the distribution function of the conductance
is mainly
 Gaussian \cite{Alt:91} 
 in a weakly disordered {\it open} sample connected to a reservoir by
broad external contacts of width
$w > \ell$, where $\ell$ is the elastic mean free path.
 Inelastic scattering processes in the reservoir result in a level broadening 
 $\gamma \sim  E_c$.
 
In the present paper the aim is to determine the distribution function 
of a dirty quantum dot with two point contacts (which allow a single 
transport channel).
 We will describe the regime of continuous energy levels, 
$\gamma \agt \Delta$, where $\gamma$ is the broadening due to inelastic 
scattering in the dot and $\Delta$ is the mean level spacing.
 This overlaps with the supersymmetric (SUSY) calculations \cite{Pri:93} 
in the ergodic regime, $\Delta \alt \gamma < E_c$. While the SUSY 
approach is valid also in the quantum regime, $\gamma <\Delta$, a 
perturbative diagrammatic approach applied below can be used also
for $\gamma\agt E_c$. In this case all diffusion modes contribute 
to the conductance rather than a single homogeneous ``zero'' mode which
is the only mode taken into account within the nonperturbative
SUSY calculations. The conductance
distribution function has a non-Gaussian shape also for such a strong level 
broadening so that the non-Gaussian shape is due to
geometric factors -- namely, the
point-like structure of contacts, rather than due to the dominance of the 
zero mode.
 In addition, we will use a standard renormalisation group technique
to consider the r\^ole of increasing disorder in the dot.

Traditionally the conductance of a system with broad, spatially homogeneous 
contacts is considered by means of the Kubo formula \cite{Kub:57}.
 For a lead geometry which involves spatially inhomogeneous currents, 
however, the conductance is often more conveniently expressed via 
scattering probabilities using the Landauer-B\"uttiker formula \cite{Lan:70}.
 In this paper, we start by writing the conductance 
in terms of Green's functions with the help of
 the Landauer-B\"uttiker formula. 
Then we will determine the conductance distribution in the case of a
continuous energy levels spectrum, $\gamma \agt \Delta$, finding
the moments of conductance by 
diagrammatic perturbation expansion in the parameter 
$( \gamma / \Delta )^{-1}$.
 Finally we will use an effective functional of the non-linear 
$\sigma$ model as a framework for the 
renormalisation group analysis necessary to describe 
dependence of the moments of conductance (and 
thus of the distribution) on the disorder parameter.  
 Following Ref.\ \cite{Alt:91}, 
we show that the $n$th order moments are proportional for large $n$ to 
exp$(un^2)$ ($u$ is a certain parameter to be specified later) which is
characteristic of a distribution function having lognormal tails. 
As the amount of disorder is increased towards the Anderson transition,
the conductance distribution in the dot becomes almost entirely lognormal.
Again, this is different from the  conductance distribution in the ensemble
of samples with broad leads which is also characterised by lognormal tails
whose r\^ole is increasing with disorder,  but remains mainly Gaussian within
the whole range of validity of the renormalisation group analysis, even at
the threshold of the transition.

We consider weak coupling through the contacts from the disordered 
region to electron reservoirs and we label probabilities for 
tunnelling through the contacts as $\alpha_1$ and $\alpha_2$ which 
are assumed to be constant.
 Following \cite{Pri:93} the level broadening due to inelastic 
scattering $\alpha$ (in units of the mean level spacing $\Delta$) 
is chosen for convenience to be greater than $\alpha_1$ and $\alpha_2$.
 So we can write $\alpha = 2\pi^2\gamma / \Delta$ where $\gamma$ is the 
total level broadening.
We will present a perturbative calculation both in the many mode regime, 
$\gamma \agt  E_c$, where the relevant small parameter is $g^{-1} = 
(2\pi^2 E_c/ \Delta )^{-1}$ ($g$ is the average
conductance of an open sample in 
units of $e^2/\pi h$) and in the zero mode regime, $\Delta \alt \gamma <  
E_c$, where the relevant small parameter is $\alpha^{-1}$.
 The calculation is not valid in the region $\alpha < 1$ which was the main 
area of interest for previous zero mode calculations \cite{Pri:93,Jal:94}.
 Since we are modelling a disordered sample, and $\alpha \agt 1$ (the level 
spectrum is continuous), we do not include Coulomb blockade and 
electron-electron interaction effects.
 As a result, the calculations are applicable to a disordered sample 
whose electronic charging energy is negligible compared to the Thouless 
energy ({\it i.e.} with spatial dimensions larger than usual quantum dots).
 However the calculations are also applicable to quantum dots when the 
gate voltage is such that the addition of a single electron does not change 
the total energy, conduction occurs and disorder effects are relevant 
\cite{Fal:94}.
 Note that a similar non-Gaussian distribution of Coulomb blockade peak height
fluctuations was found by Jalabert {\it et al} \cite{Jal:92}, and recent
experiments \cite{Cha:96} appear to be consistent with this prediction.

The framework for determination of the conductance is the 
Landauer-B\"{u}ttiker formula \cite{Lan:70}.
 We use it in the following form \cite{F+L:81}:
\begin{equation}
G = {e^2\over{2h}} \sum_{ab} \left(T_{ab}^{L} + T_{ab}^{R}\right), \label{lb}
\end{equation}
where the transmission coefficient $T_{ab}^{L(R)}$ is the probability of 
transmission from the channel labelled by $a$ in the left (right) lead to 
the channel labelled by $b$ in the right (left) lead.
 The conductance has been written explicitly in terms of transmission from 
the left {\it and} from the right since this most symmetric form is required 
to consider the influence of broken time reversal symmetry on the 
fluctuations of the conductance.
 The transmission coefficients may be related to Green's functions by 
\cite{F+L:81,Fen:88}
\begin{equation}
T_{ab}^{L(R)} = \frac{1}{(h\nu_0)^2} {\cal G}^{+(-)}\left( 
{\bf  r_1} , {\bf  r_2} ; \varepsilon \right) 
{\cal G}^{-(+)}( {\bf  r_2},  {\bf  r_1} ; \varepsilon ), 
\label{tab}
\end{equation}
where ${\cal G}^{+}$ (${\cal G}^-$) is a retarded (advanced) Green's 
function and $ {\bf  r_1}$, $ {\bf  r_2}$ are the positions of 
the point contacts.
 In the entire energy interval of interest, $\varepsilon - 
\varepsilon_F \alt \hbar /\tau$, the mean density of states $\nu_0$ 
is a constant and the $T_{ab}^{L(R)}$ are energy independent so we will 
subsequently drop the $\varepsilon$ label ($\tau = \ell /v_F$ is the 
elastic scattering time).

Each point contact has a width $w \approx \hbar k_F^{-1}$ which corresponds 
to a single channel only so that the point to point conductance, $G$, is 
obtained from the Landauer-B\"uttiker formula, Eq.\ (\ref{lb}), with only one 
term in the summation.
 Ensemble averaged cumulants of the conductance, $\aabt{G^n}$, are given
 by
\begin{equation}
\aabt{G^n} = \left[{e^2\over{h}} {\alpha_{1}\alpha_{2}\over{
(h\nu_0)^2}}\right]^n \aabt{ \left[ \gplus{ {\bf  r_1} , 
{\bf  r_2}} \gminus{ {\bf  r_2},  {\bf  r_1}} + \gplus{ 
{\bf  r_2},  {\bf  r_1}} \gminus{ {\bf  r_1} , {\bf  r_2}} 
\right]^n }.  \label{lb2}
\end{equation}
 We consider the point contacts to be separated by a distance greater 
than the mean free path so that $|{ {\bf  r_1} - {\bf  r_2}}| \gg 
\ell$.
 In Eq.\ (\ref{lb2}) spin is explicitly included with an extra prefactor of
 $2$ and the term $\alpha_{1}\alpha_{2}$ represents the transmission 
probability through the contacts themselves.
Ensemble averaging in Eq.\ (\ref{lb2}) is performed within the impurity 
diagram technique \cite{Abr:65} and it is convenient to use a representation 
in which slow diffusion modes are explicitly separated from fast 
``ballistic'' ones \cite{Gor:79}.
 Fig.~1b shows the dominant contribution to the mean conductance 
which contains one diffusion propagator (drawn as a wavy line which
 corresponds to a ladder series in the conventional technique \cite{Abr:65}).
 At each end of the diffusion propagator there are two-sided `petal' shapes
 which represent motion at ballistic scales since the average Green's 
functions $\abt{{\cal G}^{R,A}( {\bf  r} , {\bf  r^{\prime}})}$ 
(drawn as edges of the petal) decay like $\exp (-| {\bf  r} -  
{\bf  r^{\prime}} | /2\ell )$.
 At the diffusive scale, $R \agt \ell$, the petals reduce to the constant 
$\chi_2 = 2\pi\nu_0\tau$.
 The choice of diagrams is dictated by the inequality $| {\bf  r_1} -  
{\bf  r_2} | \gg \ell$.
 It means that a diagram with external points ${\bf {r_1}}$ and 
${\bf {r_2}}$ connected solely by average Green's functions as in 
Fig.~1a is exponentially small.
 
The diffusion propagator $D({\bf r},{\bf r^{\prime}};\omega)$ 
can be represented at zero frequency as
\begin{equation}
D({\bf r},{\bf r^{\prime}};0) = \frac{1}{2\pi\nu_0L^{d}\tau^2} 
\sum_{{\bf q}}  
\frac{1}{Dq^2 + \gamma} \exp [ i{\bf q}.({\bf r}-
{\bf r^{\prime}})] 
\equiv \frac{1}{2  \tau^2} \zeta(R),   \label{dr}
\end{equation}
where $D = v_F^2\tau /d$ is the diffusion constant and, for a closed system, 
the summation is carried out over all $\bbox{q} = \pi \bbox{n}/L$, where  
$\bbox{n}=(n_1,\ldots, n_d)$ are non-negative integers.
 In an open system inelastic scattering occurs in the leads and the 
summation is cut off at low momenta $q\sim L^{-1}$ where $L$ is the system 
size.
 For the system with point contacts the energy level broadening, $\gamma$, 
due to inelastic processes inside the dot is inserted `by hands' into 
Eq.\ (\ref{dr}).
 As a result, in the many mode regime, $\gamma \agt  E_c$, the summation 
in Eq.\ (\ref{dr}) may be approximated by an integration with cutoff 
at $q\sim L_{in}^{-1}$ where the inelastic scattering length $L_{in} = 
(D/\gamma )^{1/2}$.
 This leads to
\begin{mathletters}
\begin{eqnarray}
\zeta(R)= \frac{2}{(4\pi)^{d/2} \pi\nu_0 D} 
\left(\frac{2}{RL_{in}}\right)^{\epsilon/2} K_{\epsilon/2}(R/L_{in})
\approx  \cases{
  g_0^{-1}\ln(L_{in}/R),&  $d=2$\cr
(\pi /2g_0)\,(\ell/R),& $d=3 $
}.  \label{zr}
\end{eqnarray}
where $R = | {\bf  r} -  {\bf  r^{\prime}} |$, $g_0 \sim 
(\varepsilon_F \tau )^{d-1}$ is the dimensionless conductance of an open 
cube of size $\ell$, $d=2+\epsilon$ is the dimensionality of the dot, and
$K_\alpha$ is the modified Bessel function of the third kind of order 
$\alpha$.
In the zero mode regime, $\gamma \ll  E_c$, the summation in Eq.\ (\ref{dr})
 is dominated by the $q=0$ term so that
 \begin{equation}
\zeta(R) \approx {2\pi\over{\alpha}}, \qquad \alpha <  E_c/\Delta,  
\label{zeroz}
\end{equation}
\end{mathletters}
which means that the diffusion propagator is independent of $R$, 
spatial dimensionality, and the degree of disorder.

Now the leading diagram for the mean conductance, shown in Fig.~1b, 
can be evaluated. Substituting $\chi_2 = 2\pi\nu_0\tau$  for the 
petals and $\zeta(R)$ for the  diffusion propagator, one has
\begin{equation}
\abt{G} = {e^2\over{h}} \alpha_{1}\alpha_{2}\, \zeta (R), \label{meang}
\end{equation}
where $R \gg \ell$ is the separation of the point contacts.
 The mean conductance is proportional to $\zeta (R)$ and 
Eq.\ (\ref{zeroz}) shows that in the zero mode regime, 
$\gamma <  E_c$, the mean conductance depends on the 
level broadening $\gamma$, but not on the separation of the point 
contacts, the dimensionality, or the degree of disorder.
On the contrary, in the many mode regime, 
$\gamma \agt  E_c$, the mean conductance depends on 
all these parameters via Eq.\ (\ref{zr}); since it is 
inversely proportional to $g_0$ it actually increases as the amount 
of disorder increases.

In order to  calculate the variance of the conductance, we consider the 
following correlation function between the transmission coefficient from 
the single channel at $ {\bf  r_1}$ to $ {\bf  r_2}$ and the 
transmission coefficient from the single channel at $ {\bf  
{r_1}^{\prime}}$ to $ {\bf  {r_2}^{\prime}}$;
\begin{equation}
K(\Delta {\bf  r_1}, \Delta {\bf  r_2}) = 2 \left[ {e^2\over{h}} 
{\alpha_{1}\alpha_{2} \over{(2\pi\nu_0)^2}} \right]^2 \aabt{ \gplus{ 
{\bf  r_1} , {\bf  r_2}} \gminus{ {\bf  r_2} , {\bf  r_1}} 
 \gplus{ {\bf  {r_1}^{\prime}} , {\bf  {r_2}^{\prime}}} \gminus{ 
{\bf  {r_2}^{\prime}} , {\bf  {r_1}^{\prime}}}}. \label{cf}
\end{equation}
We use the notation $\Delta {\bf  r_1} =  {\bf  r_1} -  
{\bf  {r_1}^{\prime}} $, $\Delta {\bf  r_2} =  {\bf  r_2} -  
{\bf  {r_2}^{\prime}}$, where $|\Delta {\bf  r_1}|$, $|\Delta 
{\bf  r_2}| \ll \ell$. 
 The main contribution to $K$ is shown in Fig.~2a.
 It has two square ``Hikami'' boxes, $\chi_4$, representing motion at 
ballistic scales which are connected by two diffusion propagators, and 
it gives
\begin{eqnarray}
K(\Delta {\bf  r_1}, \Delta {\bf  r_2}) &=& 2 
\left[ {e^2\over{h}} {\alpha_{1}\alpha_{2} \over{(2\pi\nu_0)^2}}
 \right]^2 \chi_4 \!\left( \Delta {\bf  r_1} \right) \, \chi_4
 \!\left( \Delta {\bf  r_2} \right) \, \left[ D\left( {\bf  r_1} 
-  {\bf  r_2}\right) \right]^2,   \\
\chi_4 \!\left( \Delta {\bf r} \right) &=& \left| \int 
\gplus{{\bf k}} \gminus{{\bf k}} e^{i{\bf k.\Delta
 {\bf r} }} \, d{\bf{k}} \right|^2.
\end{eqnarray}
To determine the behaviour of $K$ on length scales $|\Delta 
{\bf  r_1}|$, $|\Delta {\bf  r_2}| \ll \ell$ we need
to evaluate $\chi_4 \!\left( \Delta {\bf r} \right)$ accurately, 
rather than substituting it by const$\times \delta (\Delta {\bf r})$ 
which is sufficient for $|\Delta {\bf  r}| \gg \ell$.
 We find that
\begin{eqnarray}
\lefteqn{\!\!\!\!\!\!\int \gplus{{\bf k}} \gminus{{\bf k}} 
e^{i{\bf k.\Delta {\bf r}}} d{\bf{k}}} \nonumber \\
 &=&  \pi\nu_0\tau \left( \frac{\pi}{2{k_{ { \! }_F}} \Delta r} 
\right)^{\epsilon/2} \!
\left[ H_{\epsilon/2}^{(1)}({k_{ { \! }_F}} \Delta r\! +\! i\Delta 
r /2\ell) +
 H_{\epsilon/2}^{(2)}({k_{ { \! }_F}} \Delta r \!-\! i\Delta r /2\ell) 
\right],
 \label{int} 
 \end{eqnarray}
where $ H_{\epsilon/2}^{(1,2)}$ are Hankel functions.
 In three dimensions this gives
\begin{equation}
K(\Delta {\bf  r_1}, \Delta {\bf  r_2}) = {1\over 2}
 \left [ {e^2 \over{h}} \alpha_{1}\alpha_{2}\zeta (R) \right]^2
\left[ {\sin^2 (k_F \Delta r_1 ) \over (k_F \Delta r_1 )^2} {\sin^2 
(k_F \Delta r_2 ) \over (k_F \Delta r_2 )^2} \, 
e^{-(\Delta r_1 +\Delta r_2 )/\ell} \right],   \label{corrk}
\end{equation}
which corresponds, in the limit $|\Delta {\bf  r_1}| = 0$, 
to the correlation function for optical speckle patterns found in 
\cite{Sha:86}.

We find from Eq.\ (\ref{int}) that $\chi_4 (0) = (2\pi\nu_0\tau)^2$ so that, 
in the single channel limit $|\Delta {\bf  r_1}| = |\Delta 
{\bf  r_2}| = 0$, we get $K(0,0) = \abt{G}^2/2$.
 This corresponds to redrawing the diagram with the external points 
$ {\bf  {r_1}^{\prime}}$ and $ {\bf  {r_2}^{\prime}}$ exactly 
equal to $ {\bf  r_1}$ and $ {\bf  r_2}$, respectively, as shown 
in Fig.~2b.
Note that although in each of the boxes in Fig.~2a
$\bbox{r}\approx\bbox{r}'$ with accuracy up to $\ell$
(as the Green's functions represented by the edges of boxes exponentially
decrease at scale $\ell$),  such an accuracy would be
insufficient for calculating the variance of the conductance of 
the point-contact dot as the area of order $\ell^{d-1}$ would include
$g_0\gg1 $ channels. 

The variance is found by ensemble averaging Eq.\ (\ref{lb2}) for $n=2$.
 Since there is symmetry arising from an overall exchange of spatial labels 
in any ensemble average, there are only two distinct contributions to the 
variance arising from the expansion of Eq.\ (\ref{lb2}).
 The first term, $[T_{ab}^{L}]^2 + [T_{ab}^{R}]^2$, is equal to the 
correlation function $K(0,0)$ from Eq.\ (\ref{cf}) contributed by the 
diagram in Fig.~2b containing two diffusion propagators.
 The second term, $2T_{ab}^{L}T_{ab}^{R}$, is contributed by a similar 
diagram containing two Cooperon propagators.
 For broken time-reversal symmetry Cooperon diagrams are absent so that, 
overall, we get
\begin{equation}
\aabt{G^2} = {1\over \beta} \abt{G}^2,  \label{vg}
\end{equation}
where $\abt{G}$ is given in Eq.\ (\ref{meang}).
 The factor $\beta$ in Eq.\ (\ref{vg}) corresponds to Dyson's orthogonal, 
unitary, and symplectic ensembles: $\beta=1$ in  the presence of potential
 scattering only, $\beta=2$ in the presence of a finite magnetic field that 
breaks time-reversal symmetry, and $\beta=4$ in the presence of weak 
spin-orbit scattering.
 For $\beta=1$ the result Eq.\ (\ref{vg}) is the well known large intensity 
fluctuations in speckle patterns \cite{Sha:86}.

In order to determine the distribution function, we need a general 
expression for the main contribution to the $n$th cumulant.
 The leading diagrams are a generalisation of those for the variance, $n=2$.
 For $n=4$, for example, Fig.~3a shows a diagram contributing to a 
correlation function between transmission coefficients from four different
 input channels to four different output channels.
 It consists of two eight-sided Hikami boxes.
 In contrast, Fig.~3b shows a contribution to the fourth cumulant 
of the single channel conductance which has two `daisy' vertices where
 each daisy consists of four petals.
 Similarly the leading diagrams for the $n$th cumulant are a generalisation
 of Fig.~3b with two daisy vertices, where each daisy consists of $n$
 petals, which are connected by $n$ diffusion or Cooperon propagators.
Each diagram gives a contribution of $\abt{G}^n$ and a factor of $(n-1)!$ 
arises because of different ways of ordering the $n$ propagators.
 For $\beta =1$ all the diagrams from the expansion of Eq.\ (\ref{lb2}) are
 present.
 However for $\beta =2$ only the diagrams without Cooperons remain
{\it i.e.} those that arise from $[T_{ab}^{L}]^n$ or $[T_{ab}^{R}]^n$.
 There are in fact only two such terms in the expansion for all values of 
$n$, whereas the total number of terms is $2^n$.
 So the relative number of $\beta =2$ terms is $2 / 2^n \equiv 
\beta^{-(n-1)}$.
 As a result, the main contribution to the $n$th cumulant is
\begin{equation}
\aabt{G^n} = {(n-1)! \over \beta^{(n-1)}} \abt{G}^n. \label{pointn}
\end{equation}
This expression corresponds to the following distribution function
\begin{equation}
f(G) = \beta^{\beta} \, {G^{\beta -1} \over \abt{G}^\beta} \, \exp \left[ 
-{\beta G \over \abt{G}} \right], \label{pd}
\end{equation}
which is drawn in Fig.~4.
For $\beta = 1$ the distribution peaks at zero conductance, whereas
for $\beta = 2$ it peaks at $\abt{G} / 2$.
 It has the  same form as the distribution
for level width fluctuations of quantum dots in the resonance regime which
was found in \cite{Jal:92} based
upon the {\it hypothesis} that chaotic dynamics in the dot are described
by random-matrix theory. The result obtained here is based upon entirely
microscopic calculations.
However, for $\beta = 2$, it disagrees with the result of
microscopic calculations by
Prigodin {\it et al} \cite{Pri:93}
 within the SUSY approach.
Their result for $\beta=2$ is the same as our $\beta = 1$ result.
 This discrepancy arises from  a different original definition
of the conductance.
Had we defined cumulants as
averages of $[T_{ab}^{L}]^n$ only, we
would have the same result as in Ref.\ \cite{Pri:93}, 
as one expects in
the region where both the exact zero-mode integration  within the SUSY approach
and straightforward diagrammatics are equally applicable. However the
conductance is defined \cite{F+L:81,Fen:88}
as the sum of $T_{ab}^{L}$ and $T_{ab}^{R}$,  Eq.\ (\ref{lb}).
When time-reversal invariance is broken by a magnetic field (i.e.\ for
the $\beta = 2$ symmetry class),  the left and
right transmission coefficients are no longer equal
for a generic asymmetric dot. Thus cross-terms like $[T^L]^m [T^R]^{n-m}$
no longer contribute to the $n$th moment of the conductance, producing the
result different from the $\beta=1$ case. It means that breaking
time-reversal invariance suppresses small amplitudes in the distribution
(\ref{pd}) and increases the mean amplitude. This has already been noted by
Jalabert {\it et al} \cite{Jal:92}  and we refer to their paper for further
discussion.

The distribution (\ref{pd}) is very simple but profoundly
different from the conductance distribution
of an open system (with broad multi-channel external contacts). In the
latter case, the variance is universal (of order $e^2/\hbar$)
\cite{Al:85,L+S:85}, and higher moments are much smaller than the variance
so that the distribution is almost Gaussian \cite{Alt:91}.  The tails
of this distribution decrease, however, much slower r than Gaussian tails.
We will show that this is
also the case for the single-channel conductance distribution considered here.
It is known \cite{Alt:91} that expressions for cumulants of the conductance 
of an open system found in the lowest order 
of perturbation theory are not valid for $n^2 \agt \zeta_0^{-1}$ where 
$\zeta_0$ is the standard weak-localisation parameter:
$\zeta_0 \equiv \zeta (R=\ell)$, Eq.\ (\ref{zr}).
 The reason is that the number of additional diagrams containing closed 
diffusion loops which describe higher order (in $\zeta_0$) contributions 
to the $n$th cumulant increase so fast that it is $n^2\zeta_0$ rather than 
$\zeta_0$ which takes the place of the effective perturbation parameter.
 We have found that corrections in powers of $\zeta_0$ also arise in the 
present case of the conductance fluctuations of a system with single channel
 contacts.
 For example one such correction which consists of diffusion propagators and 
contributes to the $2T_{ab}^{L}T_{ab}^{R}$ term of the variance is shown in 
Fig.~5.
 Three similar corrections containing Cooperon propagators also occur so 
that, for the vertex corrections in the first power of $\zeta_0$, we get 
$\aabt{G^2} = (4 /\beta^2) \abt{G}^2 \zeta_0$.
 Similarly for the $n$th cumulant extra impurity ladders can be placed 
in $n (n-1)$ different places so that the corrections give a series of 
terms in $n^2 \zeta_0$, not $\zeta_0$.
 At large enough $n$, this enhancement of corrections by a factor $n^2$
 means that the ``main'' contributions no longer dominate.

In order to find expressions for large $n$ cumulants we need to sum all 
the corrections in powers of $\zeta_0$ which is not practical within the 
diagram technique.
 Instead the summation is performed using the renormalisation group 
procedure which is carried out in the framework of an effective field 
theory, a non-linear $\sigma$ model \cite{Weg:79}, where averaging over 
realisations of disorder and averaging over {\it fast} degrees of freedom
 are performed in the derivation of the model.
 The averaging produces expressions for the $n$th cumulant of the point
 contact conductance in terms of functional derivatives with respect to 
a source field ${\sf h}( {\bf  r} )$ (for notations see \cite{Alt:91}),
\begin{equation}
\aabt{\prod_{i=1}^{n}G_{i}} = \left({2e^2\over{h}} 
{\alpha_{1}\alpha_{2}\over{(2\pi\nu_0)^2}} {\tau^2\over 8 N^2}\right)^n 
\left[ \prod_{i=1}^{n} {\rm {tr}} \left( {\delta^2 \over \delta {\sf h}_i(
 {\bf  r_1} ) \delta {\sf h}_i( {\bf  r_2} )} \right) \right] \, 
\abt{Z[ {\sf h} ]} \, \Big|_{\omega = N =0},  \label{gcum}
\end{equation}
where $\abt{Z[ {\sf h} ]}$ is a generating functional,
\begin{equation}
\abt{Z[ {\sf h} ]} =
{\int {\cal D} {\sf Q} \, \exp - F[{\sf Q};{\sf h}] \over \int 
{\cal D} {\sf Q} \,\exp - F[{\sf Q};0] },
 \qquad F[{\sf Q};{\sf h}] = F[{\sf Q}] + F_h[{\sf Q};{\sf h}] . 
 \label{zh}
\end{equation}
Here the functional $F[{\sf Q}]$ is a modification of the 
standard $\sigma$ model functional,
\begin{equation}
F[{\sf Q}] = \int  d^d r \left[ \frac{\pi\nu_0 D}{8} \mbox{Tr}  \left( \nabla 
{\sf Q}\right) ^2 d^dr
 - \frac{\pi\nu_0\gamma}{4} \mbox{Tr}\left( \Lambda {\sf Q}\right)
 \right],   \label{fstan}
\end{equation}
which takes account of the non-zero level broadening $\gamma$ (see discussion 
after Eq.\ (\ref{dr})).
The source field functional is
\begin{equation}
F_h[{\sf Q};{\sf h}] =
\sum_{m=1}^{\infty} F_h^{(m)}[{\sf Q};{\sf h}] = 
{\pi\nu_0\over 2\tau} \sum_{m=1}^{\infty}
 \Upsilon_m \, \int {\rm {Tr}} ({\sf h{\sf Q}})^m \, 
d^d r \,, 
\label{fh}
\end{equation}
with bare values of the charges $\Upsilon_m$ given by
\begin{equation}
\Upsilon_m^{(0)} = {(2m-3)!! \over m! }. \label{uo}
\end{equation}
The Hermitian matrix ${\sf Q}$ obeys the constraints
${\sf Q}^2=I$, Tr${\sf Q}=0$.
 It may be represented as ${\sf Q} = {\sf Q}_{\nu}^{\mu} \tau_{\mu}$ where 
$\tau_{\mu}$ are quarternion units and $\nu = \{ AB ; ij ; pp^{\prime}\}$ 
stands for a set of additional matrix elements.
 The replica indices $\{ AB \}$ run from $1$ to $N$ with the replica
 condition $N=0$ being applied to the final results, the loop indices
 $\{ ij \}$ label different conductances in the product Eq.\ (\ref{gcum}),
 and the indices $\{ pp^{\prime} \}$ distinguish retarded and advanced 
Green's functions.
 These indices are required to eliminate terms in the perturbative 
expansion of Eq.\ (\ref{gcum}) which do not correspond to those in the 
standard diagram technique.
 The matrix source field $h( {\bf  r} )$ is chosen to be Hermitian
 with the following $pp^{\prime}$ structure:
\begin{equation}
{\sf h} \equiv \pmatrix{0&h_{AB}^{0} \cr h_{BA}^{0}&0\cr} \otimes \tau_0 + 
\pmatrix{0&h_{AB}^{3} \cr - h_{BA}^{3}&0\cr} \otimes \tau_3\,. \label{hpp}
\end{equation}
 High gradient vertices \cite{Alt:91} are not included in the functional
 Eq.\ (\ref{fstan}): although they are involved in the renormalisation of 
the charges $\Upsilon_m$ in Eq.\ (\ref{uo}) this could produce only a change 
in preexponential factors irrelevant here.

The lowest order perturbational contribution to the $n$th cumulant arises 
from the term $(F_h^{(n)}[{\sf Q};{\sf h}])^2$ in the expansion of Eq.\
(\ref{zh}).
 The vertex $F_h^{(n)}[{\sf Q};{\sf h}]$ contains $n$ source
fields ${\sf h}( {\bf  r} )$
 and thus corresponds to a Hikami box with $n$ external points such as those
 in Fig.~2a for the variance and Fig.~3a for the fourth cumulant.
 This contribution is proportional to $(\Upsilon_n^{(0)})^2$ and does not 
reproduce the exact numerical coefficient of the diagram technique result, 
Eq.\ (\ref{pointn}), for single channel contacts which arise from daisy 
vertices with a single external point only (Fig.~2b for the variance and
 Fig.~3b for the fourth cumulant).
 The reason is that the derivation of the $\sigma$ model involves averaging 
over fast degrees of freedom so that it is insensitive to details on local 
length scales (of the order of $\hbar k_F^{-1}$).
 Nevertheless the $\sigma$ model accurately describes the behaviour of 
diffusive degrees of freedom which are the relevant ones for what follows.

The renormalisation group procedure allows effective summation of the 
higher order perturbative corrections which are logarithmic in $2d$.
 The net effect is to substitute renormalised values of the charges for
 bare ones in the expressions obtained by the perturbative expansion of
 Eq.\ (\ref{zh}) above.
 Results in higher dimensionalities can be qualitatively obtained by 
$d=2+\epsilon$ expansion.

The source field functional above, Eq.\ (\ref{fh}), is similar to the source
 field functional describing fluctuations of the density of states which is
 renormalised in \cite{Alt:91}.
 As a result of the renormalisation, the charges obey the following 
increase law,
\begin{equation}
\Upsilon_n \propto \Upsilon_n^{(0)} e^{u(n^2-n)},  \label{urg}
\end{equation}
where
\begin{equation}
u = \ln \frac{\sigma_0}{\sigma} = \ln \left( 1 - \zeta_0 \right)^{-1}
\end{equation}
In the weak disorder limit $u \approx \zeta_0 \ll 1$, whereas in 
the vicinity of the Anderson transition,
\begin{equation}
u =\cases{\epsilon \,\ln \,L/\ell,&$L\, \alt \,L_c\quad $ (a)\cr
\ln (1-g_c/g_0)^{-1},&$L\, \agt \,L_c\quad $ (b)}\,.
\end{equation}
Here $\sigma$ is the physical (renormalised) conductivity at length scale 
$L$ and $\sigma_0$ is the classical (bare) conductivity at length scale 
$\ell$.
 $L_c$ is the correlation length which diverges as $L_c\propto 
(g_0-g_c)^{-1/\epsilon }$ in the vicinity of the Anderson transition point
 \mbox{$g_0=g_c$}.

Substituting the renormalized charge in place of the bare charge in the
 leading perturbative results we get
\begin{equation}
\aabt{G^n} \sim  \abt{G}^n \, e^{2u(n^2-n)},  \qquad  n \agt u^{-1}. 
\label{highn}
\end{equation}
This is valid for cumulants with $n \agt u^{-1}$ whereas the universal 
expression, Eq.\ (\ref{pointn}), is valid for $n \alt u^{-1/2}$.
 The exponential increase law for high $n$ cumulants, Eq.\ (\ref{highn}),
 is similar to that of the local density of states \cite{Ler:88,Alt:91} 
and it leads to lognormal tails of the distribution function
\begin{equation}
f(\delta G) \sim {1\over \delta G} \exp\left[ -{1\over 8u} \ln^2 
\left({\delta G \over 4u\abt{G}}\right) \right],  \qquad  \delta G 
\agt \abt{G} /u, \label{pcln}
\end{equation}
where $\delta G = G - \abt{G}$.
 For weak disorder $u \approx \zeta_0 \ll 1$ so the main part of the
 distribution is due to the low $n$ cumulants and it has an exponential 
shape Eq.\ (\ref{pd}).
 Some very large $n$ cumulants follow Eq.\ (\ref{highn}) and the 
exponential distribution will have lognormal tails which appear for 
fluctuations $\delta G \agt \abt{G} /u$.

As the amount of disorder increases then $u$ increases in magnitude, more 
of the cumulants follow Eq.\ (\ref{highn}), and the lognormal tails become 
larger.
 Due to the condition of validity of the high cumulant expression, $n \agt 
u^{-1}$, the whole distribution will become lognormal in the region $u \sim
 1$.
 This crossover from the exponential to the lognormal distribution occurs 
before the Anderson transition {\it i.e.} still within the metallic regime,
 since $u = \ln \sigma_0 / \sigma$ then $u \sim 1$ can occur for $\sigma_0 >
 \sigma \gg 1$.
 This is similar to local density of states fluctuations
 \cite{Ler:88} where a crossover from nearly Gaussian to completely lognormal 
occurs in the metallic regime for $u \sim 1$.

Note that the lognormal distribution for local fluctuations originally 
obtained by the renormalisation group treatment \cite{Alt:91} has been 
rederived directly within the SUSY approach
\cite{M+K:95}. It is possible that the high gradient expansion 
[see note after Eq.\ (20)] corresponds to probing the new 
inhomogeneous vacuum found in \cite{M+K:95}. However the new approach
 is applicable only to the weak disorder limit, $u \approx \zeta_0$, 
and could not describe the distribution of the many channel conductance.

In summary, using diagrammatic perturbation expansion in the parameter 
$( \gamma / \Delta )^{-1}$, $\gamma \agt \Delta$, we reproduced the 
exponential distribution of conductance fluctuations in quantum dots
 with two single channel leads \cite{Pri:93} in the zero mode regime,
 $\Delta \alt \gamma <  E_c$, and we demonstrated strong dependence on
 time-reversal symmetry.
 We have shown that the distribution has the same shape in the many mode 
regime, $\gamma \agt  E_c$, but, in contrast to the zero mode regime, 
the mean and the variance are dependent on the spatial dimension, the 
degree of disorder, and the separation of the leads.
 Using the renormalisation group procedure we have shown that the exponential 
distribution has lognormal tails in both of the above regimes.
 As disorder increases, the lognormal asymptotics become more important 
and eventually there will be a crossover to a completely lognormal 
distribution.

\acknowledgments
We are grateful to Y.~Gefen, V.~E.~Kravtsov, and R.~A.~Smith for useful 
discussions.
 This work was supported by EPSRC grant GR/J35238.

\begin{figure}
\epsfxsize=0.8\textwidth
 \epsffile{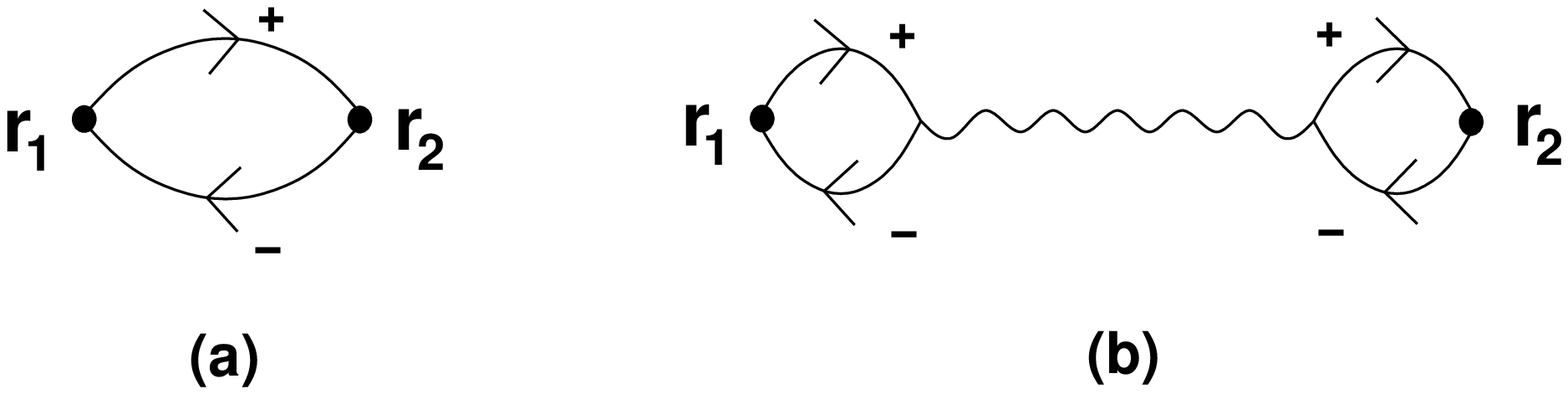}
\vspace*{.25cm}

\caption{\baselineskip=16pt 
Lowest order perturbational contributions to the mean conductance:
 (a) exponentially small diagram for $| {\bf  r_1} -  {\bf  r_2} | 
\gg \ell$, (b) dominant contribution.
}
\end{figure}

\begin{figure}[b]
\epsfxsize=0.8\textwidth
 \epsffile{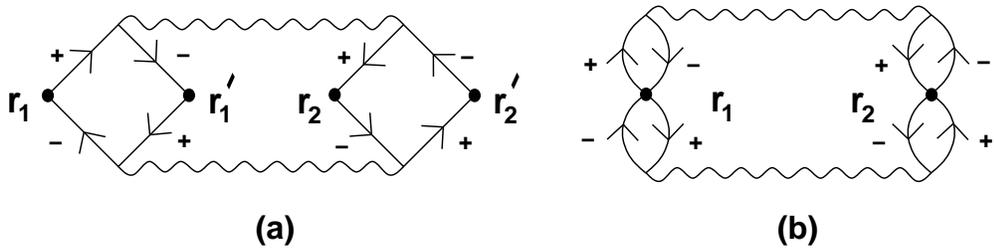}
\vspace*{.25cm}
\caption{\baselineskip=16pt 
Lowest order perturbational contributions to the correlation function 
$K(\Delta {\bf  r_1}, \Delta {\bf  r_2})$:
 (a) multi-channel correlations; 
(b) single-channel variance.
}
\end{figure}
\begin{figure}[b]
\epsfxsize=0.8\textwidth
 \epsffile{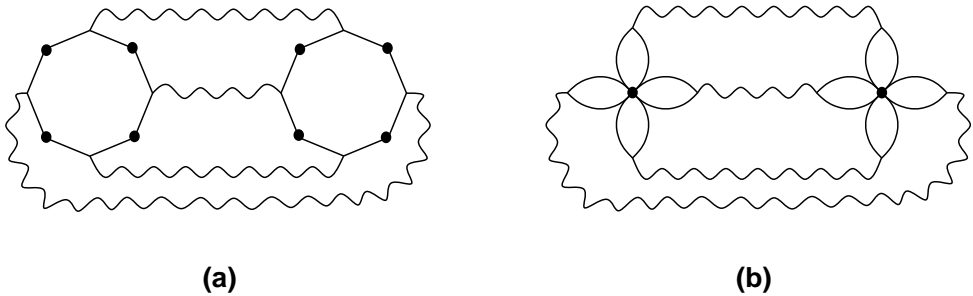}
\vspace*{.25cm}
\caption{\baselineskip=16pt
Lowest order perturbational contributions to the fourth order correlation
function.
 (a) multi-channel correlations; (b) fourth
cumulant of single channel conductance.
}
\end{figure}
\begin{figure}
\epsfxsize=\textwidth
 \epsffile{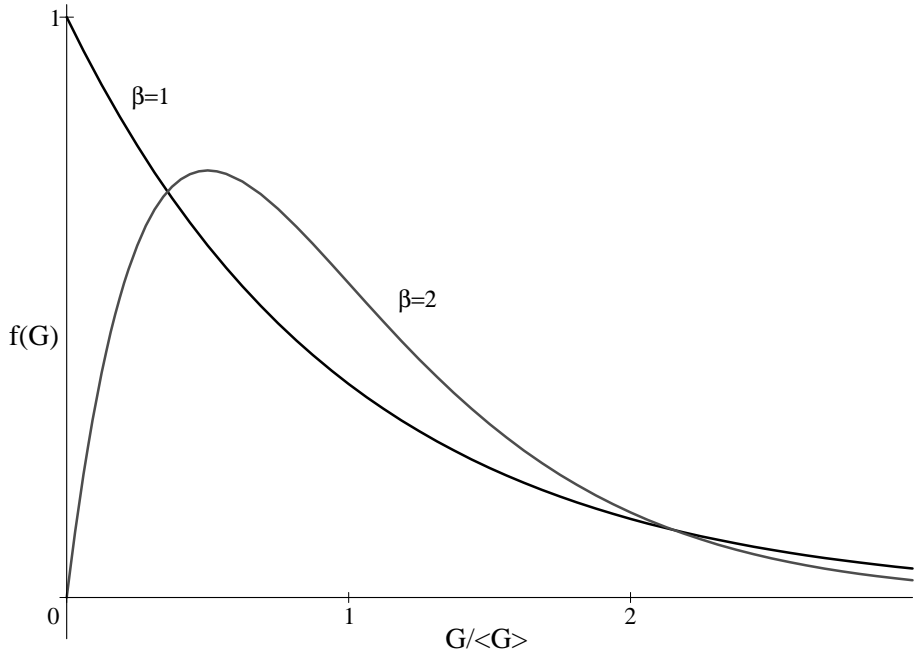}
\vspace{1cm}
\caption{\baselineskip=16pt
Point Contact Distribution Function.
}
\end{figure}
 
\begin{figure}
\epsfxsize=0.8\textwidth
 \epsffile{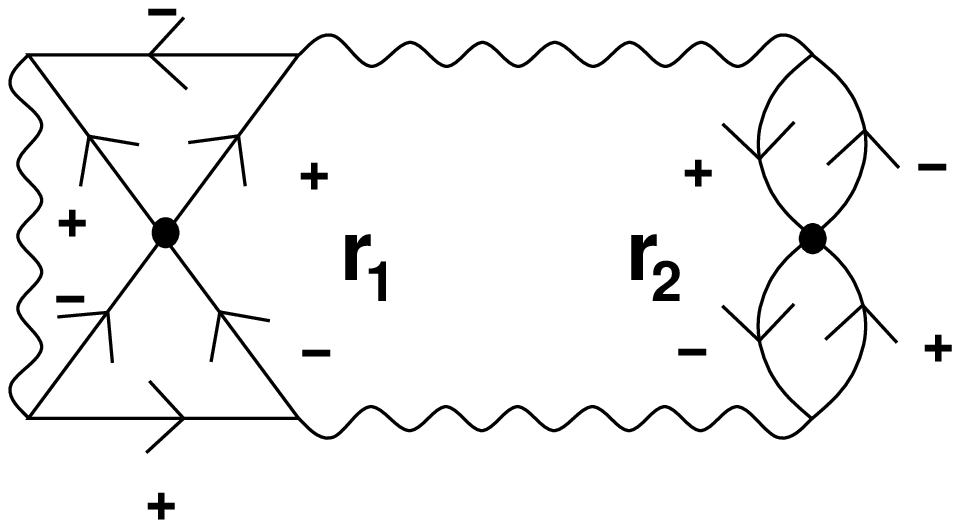}
\vspace{.25cm}
\caption{\baselineskip=16pt
Lowest order in $\zeta_0$ correction to the variance.
}
\end{figure}

\begin{thebibliography}{999}
\small\baselineskip=18pt
%
\bibitem{Pri:93} V.N. Prigodin, K.B. Efetov, and S. Iida, 
Phys. Rev. Lett.\ {\bf 71}, 1230 (1993).
\bibitem{Jal:94} R.A. Jalabert, J.-L. Pichard, and C.W.J. Beenakker, 
Europhys. Lett.\ {\bf 27}, 255 (1994).
\bibitem{Meh:91} M.L. Mehta, {\it Random Matrices} 
(Academic, New York, 1991).
\bibitem{Efe:83} K.B. Efetov, Adv. Phys.\ {\bf 32}, 53 (1983).
\bibitem{Alt:91} B.L. Altshuler, V.E. Kravtsov, and I.V. Lerner, 
{\it Mesoscopic Phenomena in Solids}, edited by B.L. Altshuler, P.A. Lee, 
and R.A. Webb, (Elsevier, Amsterdam, 1991).
\bibitem{Kub:57} R. Kubo, J. Phys. Soc. Jap.\ {\bf 12}, 570 (1957).
\bibitem{Lan:70} R. Landauer, Philos. Mag.\ {\bf 21}, 863 (1970); 
M. B\"uttiker, Phys. Rev. Lett.\ {\bf 57}, 1761 (1986).
\bibitem{Fal:94} V.I Fal'ko, and K.B. Efetov, Phys. Rev. B\ {\bf 50},
 11267 (1994).
\bibitem{Jal:92} R.A. Jalabert, A.D. Stone, and Y. Alhassid, 
Phys. Rev. Lett.\ {\bf 68}, 3468 (1992).
\bibitem{Cha:96} A.M. Chang, H.U. Baranger, L.N. Pfeiffer, K.W. West, 
and T.Y. Chang, Phys. Rev. Lett.\ {\bf 76}, 1695 (1996); 
J.A. Folk, S.R. Patel, S.F. Godijn, A.G. Huibers, S.M. Cronenwett,
C.M. Marcus, K. Campman, and A.C. Gossard, 
Phys. Rev. Lett.\ {\bf 76}, 1699 (1996).
\bibitem{F+L:81} D.S. Fisher, and P.A. Lee, Phys. Rev. B\ {\bf 23}, 6851
 (1981).
\bibitem{Fen:88} S. Feng, C. Kane, P.A. Lee, and A.D. Stone, Phys. Rev. 
Lett.\ {\bf 61}, 834 (1988).
\bibitem{Abr:65} A.A. Abrikosov, L.P. Gor'kov, and I.Y. Dzyaloshinskii, 
{\it Quantum Field Theoretical Methods in Statistical Physics} 
(Permagon, Oxford, 1965).
\bibitem{Gor:79} L.P. Gor'kov, A.I. Larkin, and D.E. Khmel'nitskii,
JETP Lett.\ {\bf 30}, 229 (1979); 
S. Hikami, Phys. Rev. B\ {\bf 24}, 2671 (1981).
\bibitem{Sha:86} B. Shapiro, Phys. Rev. Lett.\ {\bf 57}, 2168 (1986).
\bibitem{Al:85}
B.~L. Altshuler, {JETP Lett.} {\bf 41}, 648 (1985).
\bibitem{L+S:85}
P.~A. Lee and A.~D. Stone, Phys.\ Rev.\ Lett.\ {\bf 55}, 1622 (1985).
\bibitem{Weg:79} F.Wegner, Z. Phys. B {\bf 35} 207 (1979);
 K.B. Efetov, A.I. Larkin, and D.E. Kheml'nitskii,
Sov. Phys. JETP\ {\bf 52}, 568 (1980).
\bibitem{Ler:88} I.V. Lerner, Phys. Lett. A\ {\bf 133}, 253 (1988).
\bibitem{M+K:95} B.A. Muzykantskii, and D.E. Khmelnitskii, 
Phys. Rev. B\ {\bf 51}, 5480 (1995). K.B. Efetov, and V.I Fal'ko, 
Phys. Rev. B\ {\bf 52}, 17413 (1995). 



\end{thebibliography}
\end{document}